\documentclass[final
  ]
  {aipproc}

\layoutstyle{6x9}

\begin{document}

\title{PHENIX Measurement of Parity-Violating Single Spin Asymmetry in $W$ Production 
in $p$+$p$ Collisions at 500 GeV}


\author{Stephen Pate (for the PHENIX Collaboration)}{
  address={Physics Department, New Mexico State University, Las
    Cruces NM 88003, USA}
}

\begin{abstract}
The flavor-separated polarized parton distribution functions for light quarks
and anti-quarks in the proton can be studied in the production of $W$ bosons 
in $p$+$p$ collisions.  The $W$s are produced in processes like 
$u+\bar{d}\rightarrow W^+$ and $\bar{u}+ d \rightarrow W^-$ and we observe
the lepton (an electron or muon) from the decay channel
$W^{\pm}\rightarrow l^{\pm}\nu$.  The electron energy spectrum from $W$ decays 
measured with an integrated luminosity of approximately 10 pb$^{-1}$ will be shown,
with a measurement of the electron single spin asymmetry in central rapidity.
\end{abstract}

\maketitle


Over many years, the worldwide program of form factor measurements,
inclusive deep-inelastic scattering and semi-inclusive deep-inelastic scattering
experiments has made possible a definitive measurement of the polarized parton
distribution functions (PDFs) of the $u$ and $d$ quarks.  The recent global
fit by de Florian, Sassot, Stratmann and
Vogelsang~(see Reference \cite{deFlorian:2009vb} and experimental
references therein) shows a clear determination of the polarized
PDFs $\Delta u(x)$ and $\Delta d(x)$.  The polarized PDF for gluons,
$\Delta g(x)$ has been greatly constrained by the recent measurements in
$\vec{p}\vec{p}$ collisions at RHIC, but considerable uncertainties
remain in the low-$x$ region.  The polarized PDFs for the sea quarks 
($\Delta\bar{u}$, $\Delta\bar{d}$, $\Delta s$, and $\Delta\bar{s}$)
remain relatively poorly determined, on the other hand.

The PHENIX Experiment at the Relativistic Heavy Ion Collider (RHIC) in
Brookhaven National Laboratory is measuring parity-violating
longitudinal single spin asymmetries in $W$ production in 
$\vec{p}\vec{p}$ collisions that are sensitive to the light quark
sea contribution to the proton spin.  These asymmetries arise due
to the fixed-helicity couplings in the production of the $W$, i.e.
$u_L \bar{d}_R \rightarrow W^+$ and $d_L\bar{u}_R \rightarrow W^-$.
The $W$ may then be observed via its leptonic decay 
$W^{\pm}\rightarrow l^{\pm}\nu$ which produces a high-$p_T$ lepton
which is detected.  The single-spin asymmetry in the number of
observed leptons is~\cite{deFlorian:2009vb,Nadolsky:2003ga}
$$A_L^{l^+} = \frac{\Delta\bar{d}(x_1)u(x_2)(1+\cos\theta)^2 
- \Delta u(x_1)\bar{d}(x_2)(1-\cos\theta)^2}
   { \bar{d}(x_1)u(x_2)(1+\cos\theta)^2 
+ u(x_1)\bar{d}(x_2)(1-\cos\theta)^2}$$
$$A_L^{l^-} = \frac{\Delta\bar{u}(x_1)d(x_2)(1-\cos\theta)^2 
- \Delta d(x_1)\bar{u}(x_2)(1+\cos\theta)^2}
{ \bar{u}(x_1)d(x_2)(1-\cos\theta)^2 
+ d(x_1)\bar{u}(x_2)(1+\cos\theta)^2}$$
where $\theta$ is the lepton decay angle in the partonic
center-of-mass system, and $x_{1,2}\equiv (Q/\sqrt{s})e^{\pm y_W}$.
In certain kinematic limits, this asymmetry can have a very simple
interpretation;  for negative leptons detected at large forward (backward)
rapidity, the asymmetry $A_L^{l^-}$ is very nearly equal to 
$\Delta d/d$ ($\Delta\bar{u}/\bar{u}$).  In general, the
measurement provides a linear combination of polarized parton
distributions functions which must be combined with other measurements
to provide a flavor separation.

In the 2009 Run at RHIC there were polarized $pp$ collisions
at $\sqrt{s}=500$ GeV for physics for the first time.  Over a four
week period, concurrent machine development and physics
data-taking permitted significant improvements in delivered
luminosity, an in-depth study of polarization transmission during
the energy ramp up to $\sqrt{s}=500$ GeV, and a first look at
$W$ production as a tool for studying the proton spin.  The integrated
luminosity at PHENIX, including the effect of a cut on the
longitudinal distribution of collision vertices, was 8.6 pb$^{-1}$.
The average polarization of the beams was $0.39\pm 0.04$.
The beam polarization was monitored by a combination of
high-rate proton-carbon scattering events from an unpolarized
carbon target that provide a relative
measurement of proton polarization during the run, low-rate
proton-proton scattering events from a polarized hydrogen gas jet
that provide an absolute calibration of the beam polarization, and
a local polarimeter at PHENIX that makes use of the transverse
single spin asymmetry of forward neutrons observed in the
zero-degree calorimeter (ZDC)\footnote{See the contribution by
Sebastian White.}.

\begin{figure}[ht]
 \includegraphics[height=.35\textheight]{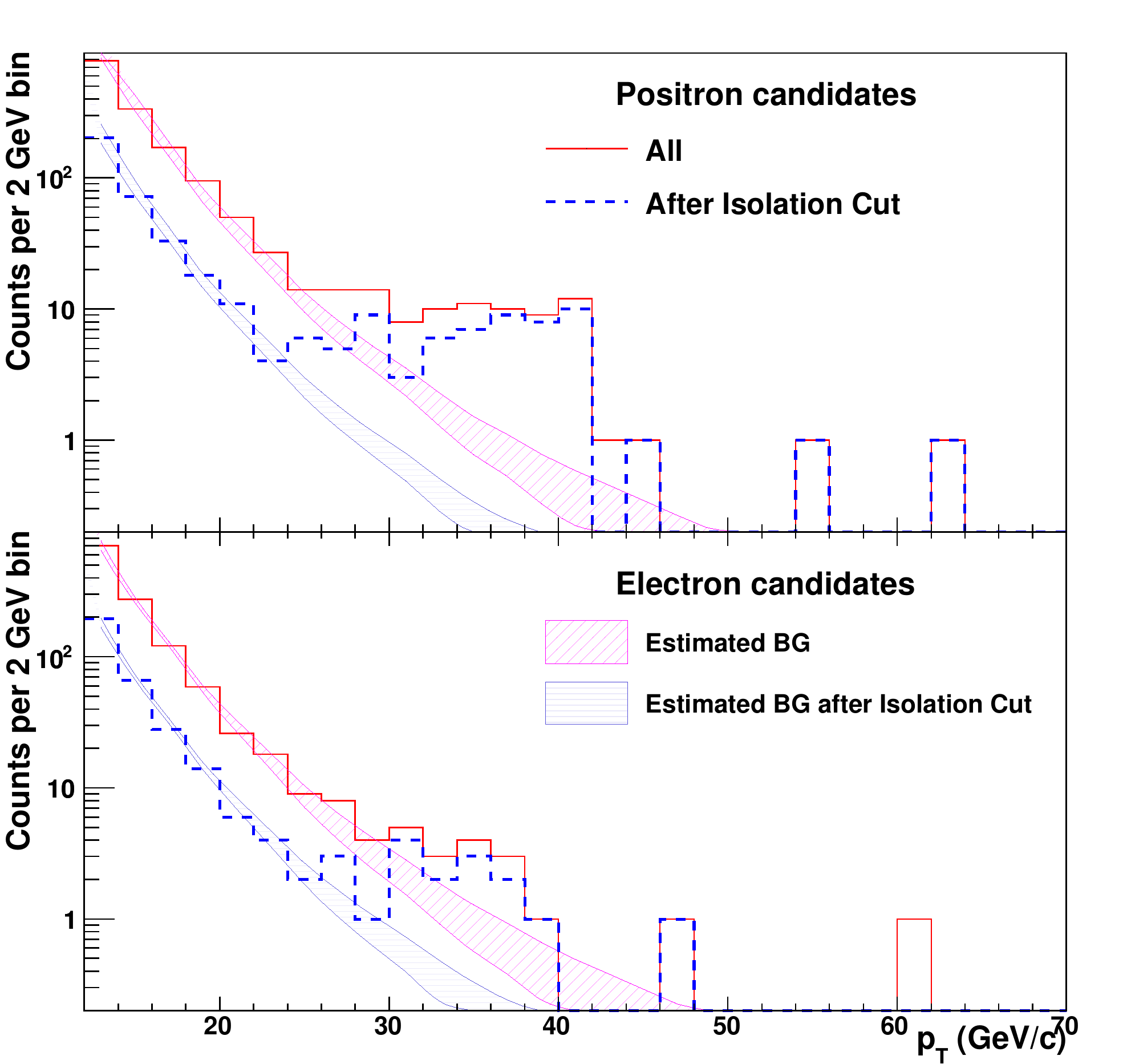}
\label{fig1}
 \caption{The spectra of positive (upper panel)
and negative (lower panel) candidates before (solid histogram)
and after (dashed histogram) an isolation cut. The estimated
background bands are also shown. The computation of the
background before the isolation cut contains contributions
primarily from photon conversions before the drift chamber and from
charged hadrons; this is described in Ref~\cite{Adare:2010xa}.
The background band after the isolation cut is computed
by scaling the background before the isolation cut by the
isolation cut efficiency measured in the background region
(12 < $p_T$ < 20 GeV/$c$).}
\end{figure}

During the 2009 Run, only the central arms of PHENIX were prepared
to observe the very low rate of very high $p_T$ leptons that arise
from $W$ decay.  (In future runs at 500 GeV, after ongoing upgrades,
also the PHENIX muon
detector arms at forward and backward rapidity will be able to 
observe muons from $W$ decay with acceptable backgrounds.)
The PHENIX central arms\footnote{For PHENIX overview, see the contribution
by Murad Sarsour.} cover $|\eta|<0.35$ in rapidity, and
the two arms cover a total of $\pi$ radians in azimuth.  For the
purposes of observing these high-$p_T$ electrons and positrons,
PHENIX employed a drift chamber and pad chambers to measure
the deflection of the trajectory of the particle
in the PHENIX central magnet, and an electromagnetic
calorimeter (of both PbGl and PbSc construction) to measure the
energy of the particle.  A variety of cuts (matching particle tracks to
calorimeter clusters, a timing cut to eliminate cosmic ray events, 
and an $E/p$ cut to eliminate hadrons) reduced the backgrounds from
QCD processes (mostly pion production) to a level where a Jacobian
signal in the electron and positron $p_T$ spectra could be seen; see
solid red histogram in Figure~\ref{fig1}.
For the purposes of determining the longitudinal single-spin
asymmetry, an additional isolation cut was made -- we required
that that total amount of additional energy and momenta
in a cone of radius 0.5 in $\eta$ and $\phi$
around the identified cluster was less than 2 GeV.  This last cut
reduces the background in the signal region ($30 < p_T < 50$ GeV/$c$)
by a factor of about 4;
see dashed blue histogram in Figure~\ref{fig1}.
We are not able to exclude events due to $Z$ production and decay from
our data sample; the number of $Z$ events expected is
small, about 7\% of the $W^+$ sample and
about 30\% of the $W^-$ sample.

\begin{figure}[ht]
 \includegraphics[height=.35\textheight]{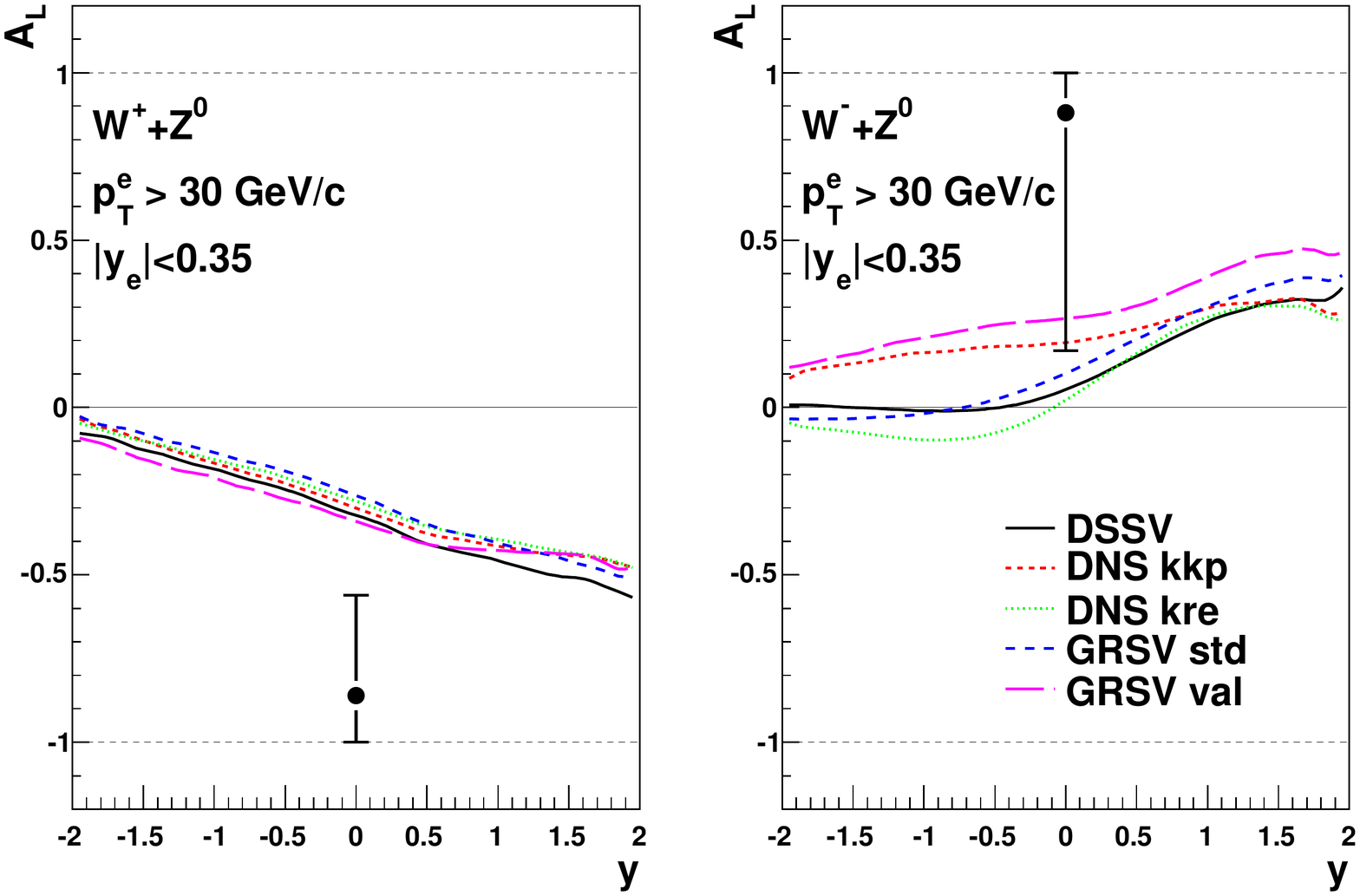}
\label{fig2}
  \caption{Longitudinal single spin asymmetries
for electrons and positrons from W and Z decays. The error
bars represent 68\% CL. The theoretical curves are calculated
using different polarized PDFs~\cite{deFlorian:2010aa}.}
\end{figure}

The measured experimental single-spin
asymmetry was determined from the experimental data by combining
the number of high-$p_T$ leptons observed when one of the beams
had positive (right-handed) helicity, $N^+$, with the number observed
when one of the beams had negative (left-handed) helicity, $N^-$:
$$\epsilon_L = \frac{N^+ -R\cdot N^-}{N^+ +R\cdot N^-}$$
where $R$ is the relative integrated luminosity of the positive to 
the negative helicity collisions, $L^+/L^-$.  The physical
parity-violating
single-spin asymmetry was then determined using the polarization $P$
of the beams and correcting for the dilution $D$ of the asymmetry due
to background events:
$$A_L = \frac{\epsilon_L\cdot D}{P}.$$
In order to maximize the usefulness of the rather limited statistics
in this initial measurement (42 positron events and 13 electron
events) a likelihood function was used to determine confidence
intervals for $A_L$; this allowed us to make use of the constraint
that $A_L$ must lie in the range $-1$ to $+1$.  The results of this 
analysis are shown in Table~\ref{table1}.  Even in this first
measurement, the results are striking.  This is the first observation
of parity-violation in the production of $W$ bosons; $A_L^{e^+}$ is
non-zero at 95\% confidence level.  In Figure~\ref{fig2} these results
are shown in comparison to predictions~\cite{deFlorian:2010aa} 
of these asymmetries based on current fits of polarized PDFs.
These results have been submitted for publication by
the PHENIX collaboration~\cite{Adare:2010xa}.  The
STAR collaboration at RHIC has made a measurement
of $W$ production as well~\cite{Aggarwal:2010vc}.

\begin{table}
\begin{tabular}{ccccc}
\hline
Sample & $\epsilon_L$ & $A_L^e (W+Z)$ & 68\% CL & 95 \% CL \\
\hline
Bkgrnd $+$ & $-0.015 \pm 0.04$ & & & \\
Signal $+$ & $-0.31 \pm 0.10$ & $-0.86$ & [$-1$, $-0.56$] & [$-1$, $-0.16$] \\
\hline
Bkgrnd $-$ & $-0.025 \pm 0.04$ & & & \\
Signal $-$ & $ 0.29 \pm 0.20$ & $+0.88$ & [$0.17$, $1$] & [$-0.60$, $1$] \\
\hline
\end{tabular}
\caption{Parity-violating single-spin asymmetry $A_L$.}
\label{table1}
\end{table}

To summarize, PHENIX has observed
a non-zero parity-violating
asymmetry in $W$ production
in polarized $pp$ collisions at 500 GeV.  These asymmetries
are sensitive to the contribution of the light quark sea
to the spin of the nucleon.  In the next few years we look
forward to the collection of 100-200 pb$^{-1}$ of additional
data on this process and providing strong constraints
in the determination of $\Delta\bar{u}(x)$ and $\Delta\bar{d}(x)$.


\begin{theacknowledgments}
  This work of this author was supported by the Office of Science of the US
  Department of Energy.
\end{theacknowledgments}



\bibliographystyle{aipproc}   

\bibliography{DIFF2010_Pate}

\begin{thebibliography}{5}
\expandafter\ifx\csname natexlab\endcsname\relax\def\natexlab#1{#1}\fi
\providecommand{\enquote}[1]{``#1''}
\expandafter\ifx\csname url\endcsname\relax
  \def\url#1{\texttt{#1}}\fi
\expandafter\ifx\csname urlprefix\endcsname\relax\def\urlprefix{URL }\fi
\providecommand{\eprint}[2][]{\url{#2}}

\bibitem[de~Florian et~al.(2009)]{deFlorian:2009vb}
D.~de~Florian, R.~Sassot, M.~Stratmann, and W.~Vogelsang, \emph{Phys. Rev.}
  \textbf{D80}, 034030 (2009), \eprint{0904.3821}.

\bibitem[Nadolsky and Yuan(2003)]{Nadolsky:2003ga}
P.~M. Nadolsky, and C.~P. Yuan, \emph{Nucl. Phys.} \textbf{B666}, 31--55
  (2003), \eprint{hep-ph/0304002}.

\bibitem[de~Florian and Vogelsang(2010)]{deFlorian:2010aa}
D.~de~Florian, and W.~Vogelsang, \emph{Phys. Rev.} \textbf{D81}, 094020 (2010),
  \eprint{1003.4533}.

\bibitem[Adare et~al.(2010)]{Adare:2010xa}
A.~Adare, et~al., \emph{submitted to Physical Review Letters}  (2010),
  \eprint{1009.0505}.

\bibitem[Aggarwal et~al.(2010)]{Aggarwal:2010vc}
M.~M. Aggarwal, et~al., \emph{submitted to Physical Review Letters}  (2010),
  \eprint{1009.0326}.

\end{thebibliography}

\IfFileExists{\jobname.bbl}{}
 {\typeout{}
  \typeout{******************************************}
  \typeout{** Please run "bibtex \jobname" to optain}
  \typeout{** the bibliography and then re-run LaTeX}
  \typeout{** twice to fix the references!}
  \typeout{******************************************}
  \typeout{}
 }

\end{document}